\documentstyle[psfig]{aa}
\def\ltsima{$\; \buildrel < \over \sim \;$}
\def\simlt{\lower.5ex\hbox{\ltsima}}
\def\gtsima{$\; \buildrel > \over \sim \;$}
\def\simgt{\lower.5ex\hbox{\gtsima}}
\def\gsimeq
{\hbox{\raise0.5ex\hbox{$>\lower1.06ex\hbox{$\kern-1.07em{\sim}$}$}}}
\def\lsimeq
{\hbox{\raise0.5ex\hbox{$<\lower1.06ex\hbox{$\kern-1.07em{\sim}$}$}}}

\begin{document}
%  \thesaurus{03(13.25.2; 11.19.1; 11.09.1: IRAS13349+2438)}

  \title{The Complex FeK line of the Narrow-Line Seyfert 1 galaxy IRAS13349+2438}

   \author{A.L. Longinotti
              \inst{1,}\inst{2},
            M. Cappi
              \inst{2},
           K. Nandra
             \inst{1},
              M. Dadina
              \inst{2}
     \and      S. Pellegrini
              \inst{3}}

   \offprints{A.L. Longinotti (all@ic.ac.uk)}

  \institute {\small 
       {(1) Astrophysics Group, Imperial College of Science, Technology and Medicine, Prince Consort Rd. SW7 2AZ, London, U.K.}\\
{(2) Istituto IASF-CNR, Sezione di Bologna Via Gobetti 101, 40129 Bologna, Italy}\\
{(3) Universita' degli studi di Bologna, Dipartimento di Astronomia, via Ranzani 1, 40127 Bologna , Italy}} 
   
   \date{Received / Accepted }

\titlerunning{XMM-Newton Observation of IRAS13349+2438}
\authorrunning{A. L. Longinotti et al.}

\abstract{
The observation of the Narrow-Line Seyfert 1 galaxy IRAS13349+2438 performed  
by XMM-Newton European Photon Imaging Camera between 0.3-10 keV, is presented here. 
The broadband spectrum of the source is dominated at low energies
(E$\leq$2 keV) by a strong excess of emission and by complex
emission/absorption features  between $\sim$ 5.5-8.0 keV.
The soft X-ray spectrum is consistent with ionized absorption found by the Reflection Grating Spectrometer (Sako et al. (2001)).
We focus on the  2-10 keV spectrum which shows
clear evidence for a broad, complex FeK$\alpha$ line, previously
unseen by ASCA, and for an Fe K-shell edge detected at $\sim$ 7.3-7.4 keV 
(rest-frame).
The presence of this edge could be explained by either
partial covering or ionized/relativistic reflection models, with the
latter being preferred and with a resulting power law slope of 
$\Gamma$$\sim$ 2.2.
The line profile is complex, with a broad bump between $\sim$ 5.5-6.5 keV and 
a narrow emission line at $\sim$ 7 keV, separated by a sharp drop at $\sim$ 6.8 
keV. 
This profile is compatible with two possible scenarios: 
i) a broad, ionized and gravitationally redshifted Laor diskline plus a narrow and ionized 
emission line; ii) a broad, ionized and single Schwarzschild (double-peaked) diskline with a 
superimposed narrow absorption line. 
}
   \maketitle

%
%  14.Sep.'90: Demo-Vs.
%________________________________________________________________

\section{Introduction}

It has been recognized that one of the most significant results 
achieved by the $ASCA$ satellite has been the discovery of 
broad fluorescent FeK lines in some Seyfert 1 galaxies 
interpreted as arising from X-ray reflection onto relativistic accretion 
disks (Tanaka et al. 1995, Fabian et al. 1995, Nandra et al. 1997, Fabian et al. 2000).

The skewed and asymmetric line profiles are consistent with that expected 
from an accretion disk orbiting a central black hole (Fabian et al. 1989; 
 Stella \& Campana 1991; Matt et al. 1993). 
The shape and strength of such a component potentially provide a powerful and 
unique diagnostic of the 
regions closest to the black hole, but a spectrum of high statistical quality 
is necessary to obtain any reliable physical conclusion.

Indeed, estimation of the line parameters depends critically on fitting the
 underlying continuum correctly, as pointed out by, e.g., Pounds \& Reeves (2002) 
who studied a small sample of Seyfert 1 observed by XMM-Newton.
In this regard, they 
emphasised the importance of analysing EPIC cameras broad-band spectra to 
identify the primary X-ray continuum because of the relatively large effective 
area above 7 keV 
and the opportunity of using the 
complementary high resolution grating spectrometers, on board both XMM-Newton 
and Chandra, to resolve and constrain superimposed emission and absorption 
features (the respective energy bands for these instruments are 0.3-2.5 keV and
0.8-7 keV).

ASCA results showed that the FeK$\alpha$ emission 
line was a common feature in Seyfert 1 galaxies and its peak energy  
was found to be $\sim$ 6.4 keV (Nandra et al. 1997).
Subsequent Chandra data have indicated that in some cases there is a separate 
narrow component superimposed on the broad emission (e.g. Yaqoob et al. 2001).
In other cases the 6.4 peak may be part 
of the disk line (Lee et al. 2002; Fabian et al. 2002).
In addition, at the time of the writing, there is a considerable uncertainty and 
discussion as to how common relativistically broadened components are in Seyfert 
galaxies (e.g. Reeves 2002, Padmanabhan \& Yaqoob 2002).
In particular no clear picture has emerged from the new data from XMM-Newton.
In some individual cases there is apparently emphatic evidence for relativistic 
effects in the FeK$\alpha$ line (e.g. Fabian et al. 2002; Turner et al. 2002),
while in many others there appears to be no clear evidence for a broad component
at all (e.g. Gondoin et al. 2001; Pounds et al. 2002).
 
IRAS13349+2438 is a powerful radio quiet quasar, with a bolometric luminosity
 $\geq$ 10$^{46}$ erg/s, at {\itshape z}=0.10764. Despite its high luminosity, this source is often 
classified as a Narrow Line Seyfert 1 (NLS1) for its striking similarities to this 
class of objects (Brandt et al. 1997a).  NLS1 galaxies  often show extremely complex 
FeK lines in XMM-Newton observations  (Mason et al. 2001; 
O'Brien et al. 2001; Boller et al. 2002).
They are thought to be characterised by high accretion rates (Pounds et al., 1995) which leads 
to the possibility that the accretion disk is ionized (Ballantyne et al. 2001).
IRAS13349+2438's  X-ray spectrum is  known to have a steep power-law ($\Gamma$$\sim$ 2.2) and a 
soft X-ray warm absorber both in ROSAT and ASCA observations 
(Brandt et al. 1996; Brandt et al. 1997b). 
The warm absorber has been strongly confirmed in the detailed XMM-Newton Reflection Grating 
Spectrometer (RGS) 0.3-2.5 keV data 
analysis of Sako and collaborators (Sako et al. 2001a). 

Here we report the XMM-Newton EPIC  observation  
of IRAS 13349+2438,  with the intent 
of exploring the broad-band (0.3-10 keV) spectral properties through the CCD cameras 
improved sensitivity above 7 keV.

\section{Observations and data reduction}
IRAS 13349+2438 was observed by XMM-Newton on the 19-20 June, 2000 during 
the Performance Verification phase, for a total duration of 62 ks. 
The data reported here come from the EPIC instrument; the pn camera was used 
in small window mode with the thin filter, while the MOS1 camera was set 
in full window with the medium filter. 
The MOS2 detector was operated in timing mode (FAST UNCOMP) and is not 
considered here.
The data were reduced using the XMM-Newton Science Analysis Software (SAS, 
version SAS\_20001203\_1730); 
the subsequent event selection has been performed taking into account the 
most recent calibrations.
We removed a period of strong contamination by soft protons flare at the end of the 
light curve, by cutting out all the 
contaminated events; taking into account also the satellite dead 
times, the filtered event files result in 44 ks of good exposure  for the MOS1 and 31 ks for 
the pn instrument.
We decided to use only pattern 0 events for pn data 
since the inclusion of higher level patterns reduces the spectral resolution of the instrument,
(in these data the pile-up percentage can be reduced  to $<$ 0.1\%  when selecting single events only); 
for MOS1 data we chose the best calibrated pattern $\leq$ 12, for which the pile-up is 
$\leq$ $\sim$ 0.01\%.

Source events have been collected from a circular region of  40$^{\prime\prime}$
of radius centred on the source, while background events have been collected from 
a similar region in a nearby source-free portion of the detectors.  
For the chosen source radius the encircled energy fraction varies between 88\% 
and 92\% of the total, from 1.5 keV to 10.5 keV.
About 93000 source counts
were collected from the pn  and 33000 from the MOS 1 detector. 
In the pn the background count rate contributes  $\sim$ 1\% to the source count rate 
between 0.4-10 keV and less than 10\% between 6-10 keV.
We checked for the effect of
spatial variations in the EPIC internal background by extracting from
multiple areas in the detector, but found no significant difference
in the final spectrum.
Pulse-height spectra were binned in order to have at least 20 counts/bin which allows
the use of  $\chi^2$ minimization for spectral fitting. 
Data from the pn camera are used in all the figures but the analysis and the 
interpretation that follows is also consistent with MOS camera data.

\begin{figure*}[tb] 
%\begin{center} 
\psfig{file=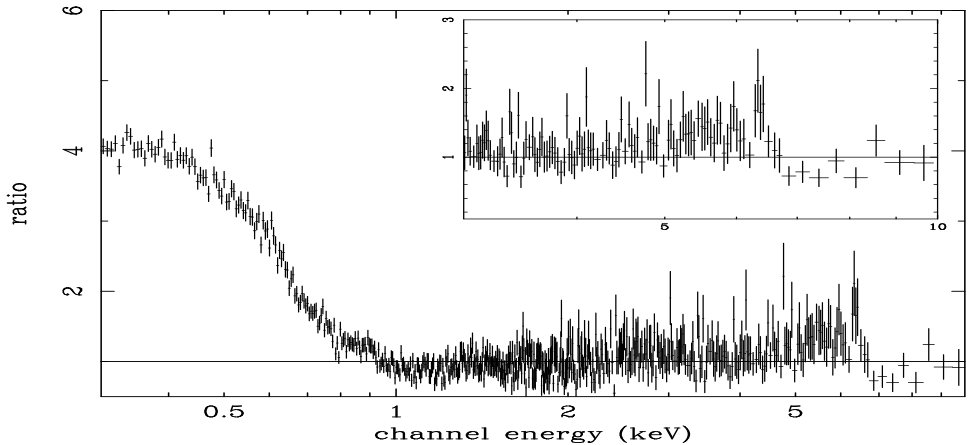,height=7cm,width=16cm,angle=-90} 
\caption{The XMM-Newton EPIC pn 0.3-10 keV spectrum of IRAS 13349+2438: the 
figure shows the data to model ratio  in the observer's frame. 
A power-law of slope 2.04 $\pm$ 0.06 has been fitted to the 2-10 keV 
data. The extrapolation of this model to lower energies data results in a clear soft 
X-ray excess. The 3-10 keV residuals in the zoom show a broad hump between 5-6 keV, a narrow 
absorption notch at $\sim$ 6.2 keV and a narrow feature peaking at $\sim$ 6.4 keV 
followed by a deep drop.} 
%\end{center} 
\end{figure*}

\section{Spectral analysis}
The source flux varied by a factor of $\sim$ 1.5  during the XMM-Newton 
observation. Time resolved hardness ratio analysis did not highlight any 
significant spectral variation, so we integrated the data over the whole exposure to 
perform the spectral analysis.

Fig. 1 shows the 0.3-10 keV spectrum plotted as the ratio of the observed 
pn data to a simple power law model fitted to the 2-10 keV data only. 
The power law slope derived is $\Gamma$= 2.04 $\pm$ 0.06.
Two striking  spectral features clearly stand out from this fit: 
\begin{itemize}
\item i) an excess of soft X-ray emission below $\sim$ 1 keV; 
\item ii) a broad feature near the Fe complex at 5-7 keV (the zoom on this feature is also 
included in Fig.1).
\end{itemize} 

{\bfseries Parametrization of the 2-10 keV spectrum} \\ 
The following basic model, 
called Model 1, aims to describe the spectral shape by parametrizing 
the data without any physical insight. 
As a crude parametrization of the Fe K$\alpha$ profile, we first used two Gaussian emission 
lines, a narrow  blue peak at a rest-frame energy of E=7.0 $\pm$ 0.1 keV, 
and a broad redshifted ``wing'' at a rest-frame energy of 
E=6.0 $^{+0.3}_{-0.2}$  keV. An absorption edge at 7.4$^{+0.3}_{-0.2}$  keV 
rest frame with optical depth of 0.44 $\pm$ 0.22,  also 
improved the fit in this model by  $\bigtriangleup \chi^2$ = 22.
The fit, with a $\chi^2$ of 208 for 216 degrees of freedom and 
$\Gamma $=2.07$^{+0.05}_{-0.15}$, gave a photon index of the 
continuum power-law slightly flatter than $\Gamma $ = 2.2, as derived from previous 
ASCA observations (Brandt et al. 1997b).
The detailed parameters of this model are listed in Table 1 while Fig. 2 shows the 
unfolded 
spectrum and the contours at the 68.3\%, 90\% and 99\% confidence levels for two 
interesting parameters of all the three components. 

{\bfseries Quick look at the 0.3-10 keV spectrum}\\
In the 0.3-10 kev  fitting, Model 1 was fixed based on the 2-10 keV 
data.   
The soft X-ray spectrum has been fitted by adding a 
blackbody component of kT = 85 $\pm$ 1 eV to the power-law. 
This gives the best description of the soft continuum  when compared to other 
continuum models tried (e.g. a single power-law, a  broken power-law and/or a 
bremsstrahlung model),  but still the residuals do require discrete 
soft X-ray components.  
We fitted these residuals by adding to the (power law + 
bbody) model either a traditional warm absorber model or a number of 
relativistic soft X-ray disklines as  recently proposed by several 
authors (Branduardi-Raymont et al. 2001, Sako et al. 2001b, Mason et al. 2002) 
for other NLSy1 type objects (very similar to IRAS13349+2438).
The relativistic emission model yields three disklines 
from O VIII, N VII and a Fe L lines blend.
These species are  
expected from ionized disks, but it is unclear whether their strengths are  consistent with 
theoretical predictions (Ballantyne et al. 2002).  
The best fit parameters of the warm absorber model, $\xi$ $\sim$ 12 erg cm s$^{-1}$, and N$_H$ $\sim$ 2.5 $\times 10^{21}$ cm$^{-2}$  are consistent with the values found by 
Sako et al. (2001b), based on RGS data (the ionization parameter is defined by $\xi$=4$\pi$ F$_x$/n$_H$). 
Both models yielded very large reductions 
in $\chi^2$. We obtained $\bigtriangleup \chi^2$ = 155 for the warm 
absorber model  and $\bigtriangleup \chi^2$ = 255 for the soft X-ray disklines 
with respect to the powerlaw+bbody $\chi^2$=846 for 562 d.o.f.
While both models are physically plausible, these spectral parametrizations  
are very much model dependent.
In reality both components maybe present and higher spectral resolution 
is required to disengtangle them.

\begin{figure*} [tb] 
%\begin{center} 
\psfig{figure=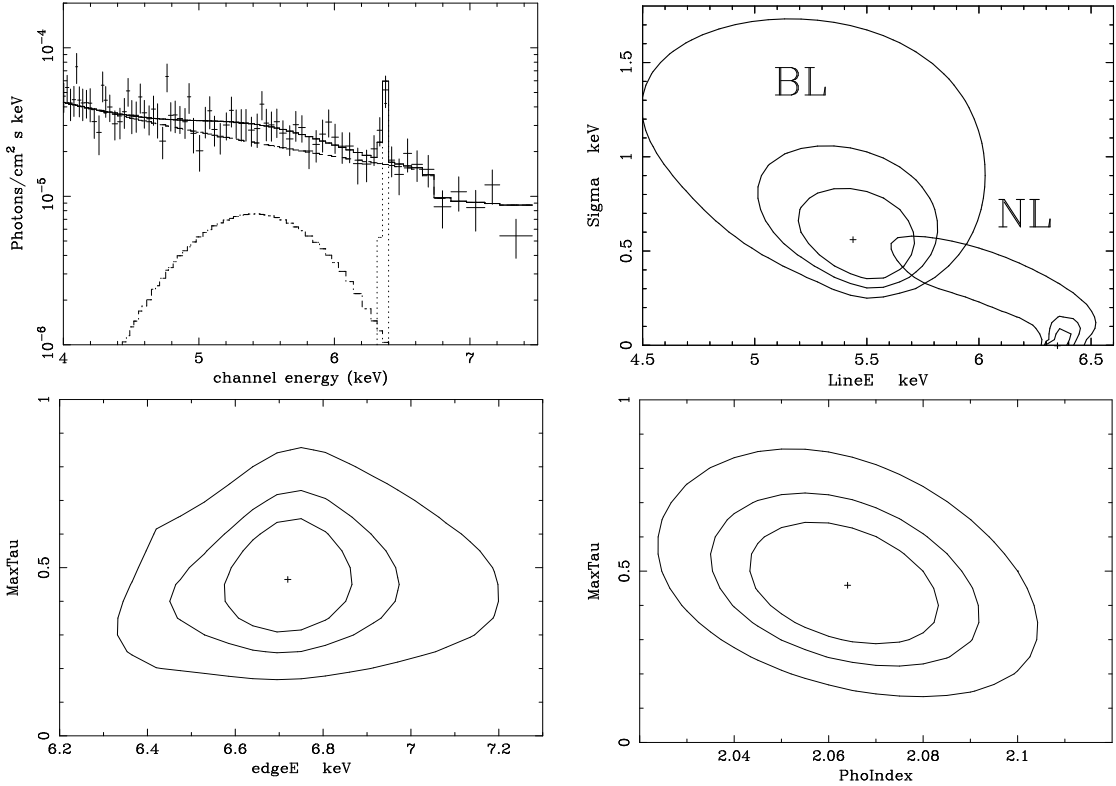,height=12cm,width=16cm,angle=-90}
\caption{Model 1: the 2-10 keV data have been parametrized by a simple 
model consisting in a power-law of $\Gamma$= 2.07$^{+0.05}_{-0.15}$, two Gaussian emission 
lines and an absorption edge; left above: 
observer's frame picture of the unfolded spectrum shows the data fitted by 
the underlying model; clockwise: contours at the 68\%, 90\% and 99\% 
confidence level for energy and width of both the broad and the narrow lines, 
for the power-law photon index and the edge optical depth,  
and for the edge energy and optical depth. All the energies in the picture are in the observer's 
frame, for a complete list of parameters, see Table 1.} 
%\end{center} 
\end{figure*}
 
In the following  sections we therefore concentrate on the hard X-ray data (above 2 keV), 
testing more physically realistic  
models than Model 1, which can in principle account simultaneously for the narrow and broad 
residuals between 5-7 keV, and for the Fe-K absorption edge.
 
\subsection{Origin of the Iron-K edge: transmission models}
The simple parametrization described before, includes a strong Fe-K
edge at $\sim$ 7.4 keV. The physical origin of this edge can be important in the
interpretation of the remainder of the spectrum (e.g. Pounds \&
Reeves 2002; Fabian et al. 2002). An origin in
line-of-sight material was first considered. This can account for the edge, but could also
cause apparent curvature in the continuum mimicking a broad disk line at lower energies.

The edge optical depth and energy in Model 1 are well constrained;
the energy of 7.4 keV implies the Iron plasma is mildly ionized (Fe X-XII).
The equivalent H-column density of the absorbing medium estimated from the 
Fe K-shell cross section and from the observed edge's $\tau$, results in 
N$_H$ $\sim$ 2.8 $\times 10^{23}$ cm$^{-2}$.
As we would then expect an Iron L-edge at $\sim$ 0.95-1.03 keV rest frame,  
we calculated the photoabsorption cross sections of K-shell and L-shell 
for that particular ionization stage, using a procedure by 
Verner et al. (1993).  
From the N$_H$  value and the L-shell cross section, the expected 
absorption L-edge optical depth is therefore estimated as $\tau$=8.6.
No edge with such optical depth is detected.
A total covering 
absorber  therefore cannot account for the FeK-edge, so the 
feasibility of a partial covering model was tested.
In general the column can be ionized so we combined the XSPEC ABSORI model (Zdziarski et al. 1990, Done et al. 1992) 
with two power laws, one unabsorbed and one subject to ionised absorption 
(reproduced by the ABSORI model).
If we are to mimic an edge with $\tau=0.4$ in
the total spectrum, this sets a minimum value for the covering
fraction. Even if the optical depth at the Fe K-edge in the absorbed
component, $\tau_{abs}$, is infinite, the edge in the combined
spectrum cannot be reproduced unless the covering fraction is greater
than $e^{-\tau_{abs}}$. We therefore set a lower limit to the
covering fraction of $C_{\rm f}>0.66$.
Such a model in principle mimics the spectrum broad 
hump and the absorption edge, reproducing the effect of a ``patchy'' absorber where
the nuclear emission leaks through a material of column density N$_H$. 
The same Gaussian narrow line of Model 1 at $\sim$ 7 keV is included in this fit to
account for the narrow peak residuals. 

The fit with Fe abundance fixed to the solar value\footnote{The abundances have been set to 
the values given by Anders \& Grevesse (1989); in particular, Fe abundance is normalized to the 
solar value of 4.68 $\times$ 10$^{-5}$  with respect to Hydrogen}  gave $\chi^2$=232/218
degrees of freedom,  with 
photon index of 2.67$^{+0.08}_{-0.11}$, column density of 
N$_H$=7.2 $\times 10^{22}$ cm$^{-2}$; 
by leaving the Fe abundance free to vary during the fit, it becomes 
greater than the solar value by a factor of 12 and the H column density decreases 
to  3.47 $\times 10^{22}$ cm$^{-2}$, with $\Gamma$=2.66 $^{+0.10}_{-0.09}$.
Fig. 3 shows this over-abundant fit, which, although still worse than Model 1, 
gave a statistical improvement ($\chi^2$ = 226/217 degrees of freedom) compared 
with the one with solar Iron abundance.
Such Fe abundance is however hard to explain.
Mildly ionized gas partially covering the X-ray source can therefore
account for the Iron edge only if it is highly overabundant in Iron, and
in any event does not account well for the broad residuals in the 5-7
keV band, where a broad emission feature is required by the data. 
An alternative origin for both line and edge is in
reflection. 
We now consider such models.
\subsection{Origin of the Iron K-edge: Reflection models}
Another possible origin for the edge feature at 7.4 keV is via
reflection in optically thick material (George \& Fabian 1991; Matt et
al. 1991). 
Indeed, any emission line produced by optically thick
material should be accompanied by a reflection continuum, as is
commonly observed  (e.g. Nandra \& Pounds 1994).
Having established the presence of reprocessing by mildly ionized gas because
of the  energies of the features in Model 1,
three different XSPEC reflection models from ionized material  have been tried for 
fitting the continuum. 
In all of them, the reflection fraction R =A$_{refl}$/A$_{pl}$ expresses the amount of 
reflection with respect to the incident X-ray irradiating flux and it is usually 
assumed to be descriptive of the source geometry 
(R=$\Omega/2\pi$, R $\sim$ 1 if the reprocessing 
material covers 2$\pi$ of the source).
We first used the model PEXRIV (Magdziarz \& Zdziarski 1995).
It turned out that the confidence region for the reflection fraction is strongly biased towards very high values of R ($\sim$ 4)  with a photon index of $\sim$ 2.2 assuming a surface temperature
 of 3$\times$ 10$^4$ K, in an optically thick  disk, with an average inclination angle of $\sim$25$^{\circ}$.
 When R is $\geq$ 2, anisotropy and/or time delay effects 
are implied, but presumably in this case the present data have a large value of R
because the reflection is forced to fit the deep
drop of the spectrum at energy greater then $\sim$ 7 keV.
Relativistic effects are very likely to be involved if the edge
originates in an accretion disk, as suggested by the broad hump in
Fig. 1 assuming this is identified with a disk line. These effects
not only broaden but shift the apparent edge threshold energy, which
may result in a poor estimate for the ionization state and optical
depth. 
To overcome these problems a smoothed edge component (SMEDGE) was added to the model  
after setting to zero the value of the Fe abundance in the PEXRIV; in this way only the SMEDGE accounts for the drop,  mimicking the relativistic smearing.
Fits with the PEXRIV/SMEDGE model
alone were, perhaps unsurprisingly, unable to account for the broad
disk-line residuals seen at 5-7 keV. We therefore added a diskline
separately. This alternative model leads to an estimation of the
reflection fraction of R $\sim$2 and $\Gamma$ $\sim$ 2.24, with an inclination for the disk of $\sim$ 30$^{\circ}$, values which are certainly more common to Seyfert 1.   
The energy of the edge is not affected very much by the type of line, being 
7.27$^{+0.19}_{-0.44}$ keV with a Schwarzschild diskline and 7.26$^{+0.15}_{-0.64}$ keV with a 
Laor diskline, see Table 1 for a complete list of parameters and $\chi^2$.
The PEXRIV/SMEDGE models have an excellent fit, depending on the line model (see below). 
Surprisingly,
more physically realistic reflection models for the edge did not
match the data particularly well. We tried both the REFSCH and XION
models, which are computations of a power law reflected on an
ionized accretion disk, convolved with a relativistic disk line, (see Magdziarz \& Zdziarski, 1995 and  Nayakshin et al. 2001), but the shape of the spectrum in 
both cases does not model the edge residuals very well and there is no statistical improvement
($\chi^2$=239/218, $\chi^2$=236/216 respectively).
\subsection{Origin of the broad 5-7 keV feature}
Neither complex absorption nor reflection appears to account well
for the broad feature between 5 and 7 keV (Fig. 1). It therefore
seems most likely that this originated as a relativistic disk line,
associated with the reflection spectrum discussed above. We tested models in
both the Schwarzschild and Kerr metrics (Fabian et al. 1989; Laor et
al. 1991). The results are shown in Table 1.
\begin{figure*}[htb]
\parbox{8truecm}
{\psfig{file=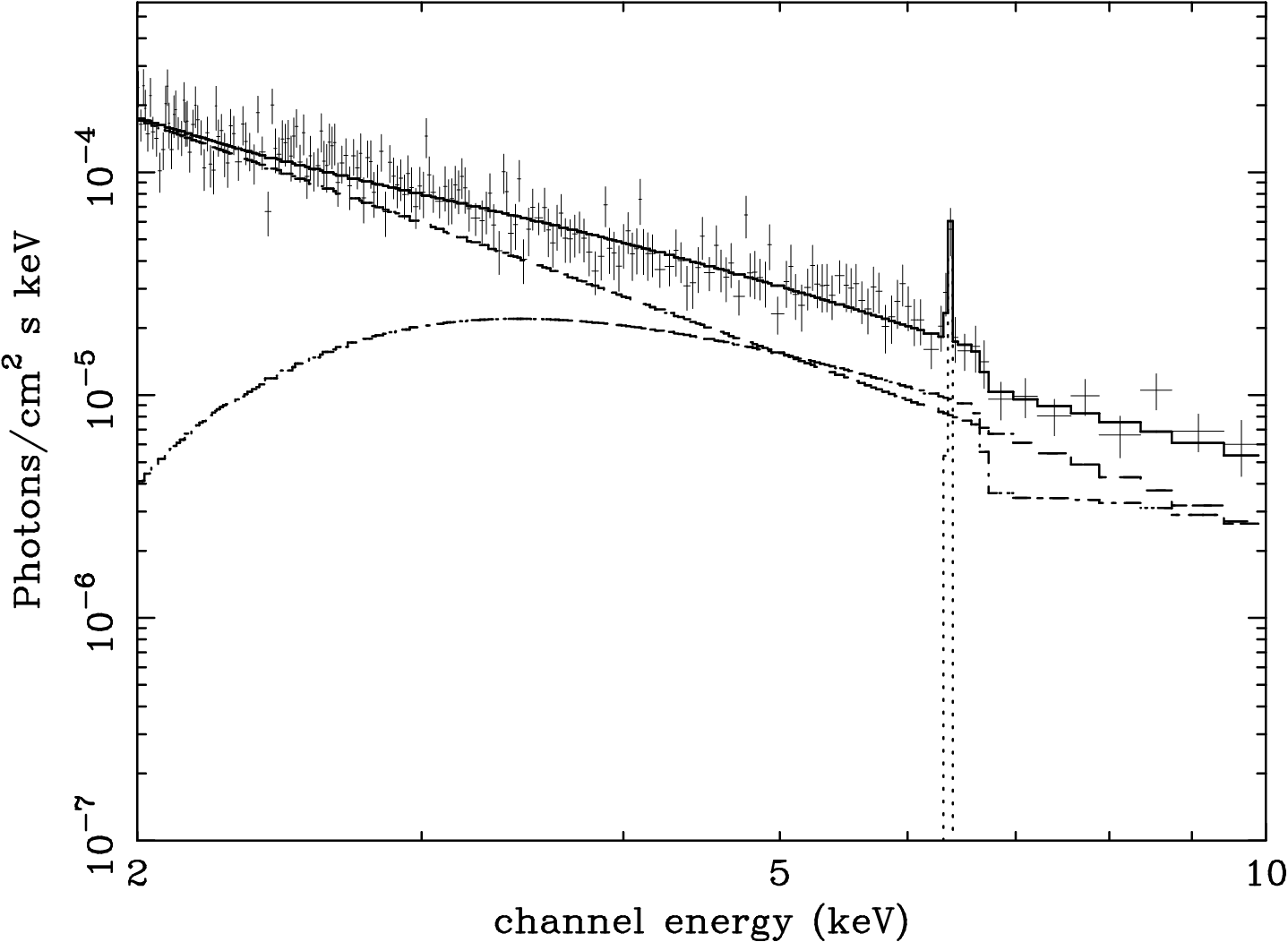,width=8.0cm,height=6cm,angle=-90}}
\  \hspace{0.1truecm}     \
\parbox{8truecm}
{\hspace{-0.5cm}\psfig{file=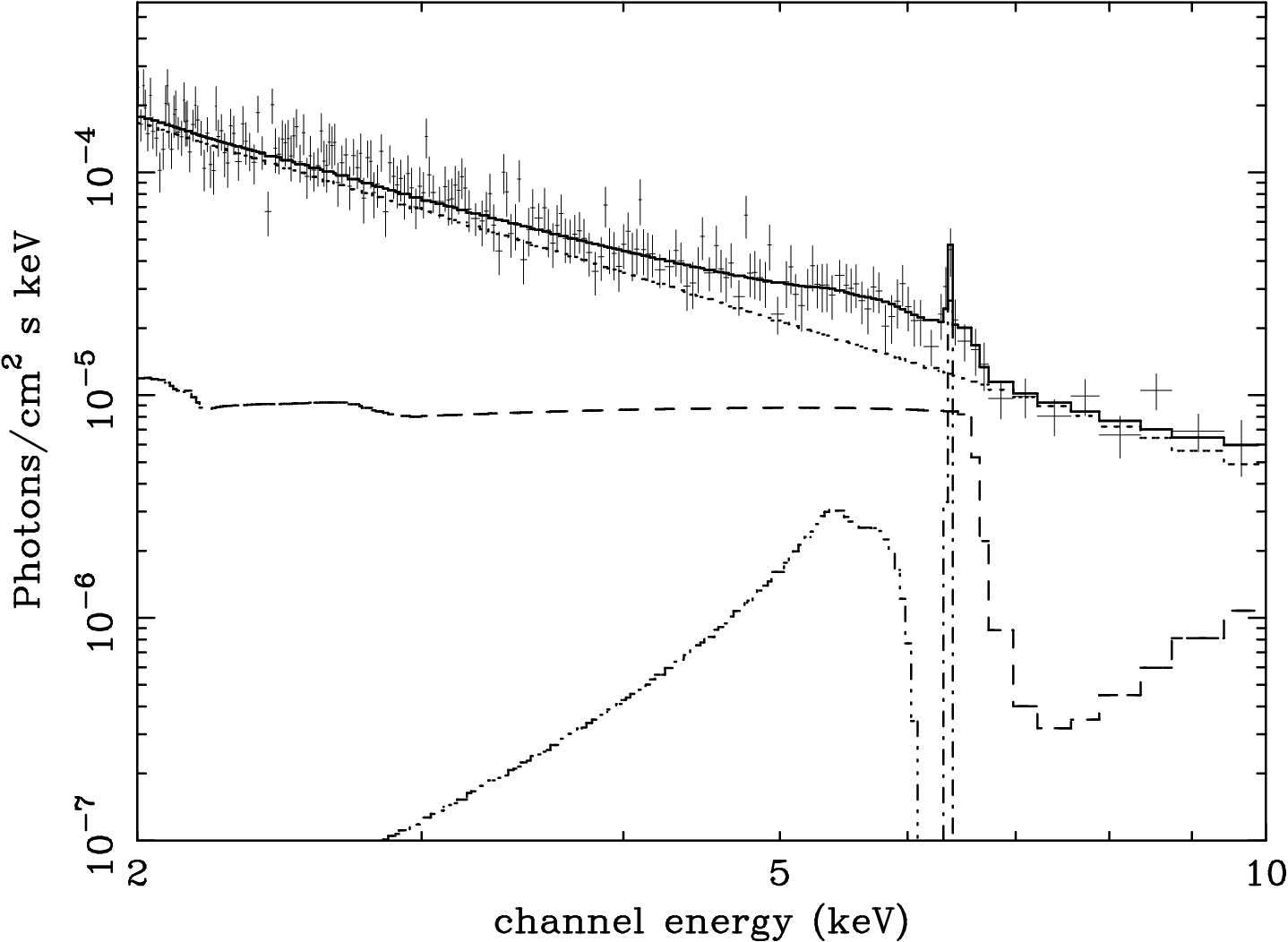,width=8.0cm,height=6.0cm,angle=-90}}
\parbox{8truecm}
{\caption{\small Observer's frame partial covering model with over abundance of 
Iron: the broad residuals at $\sim$ 5-6 keV are still clearly visible and there is no
evidence of absorption in the data below 3 keV, as required by the column 
density of a few $\times 10^{22}$ cm$^{-2}$.}}    
\  \hspace{.1truecm}     \
\parbox{8truecm}{\caption {\small  Observer's frame picture of the unfolded spectrum for model 3 of Table 1 with the FeK$\alpha$ fitted by a Kerr diskline plus a narrow Gaussian emission line 
(the rest frame energies are respectively 6.67$^{+0.72}_{-0.67}$ keV  and 7.02$^{+0.25}_{-0.06}$ keV).}} %\ %\vspace{0.4truecm} \
\parbox{8truecm}
{\psfig{file=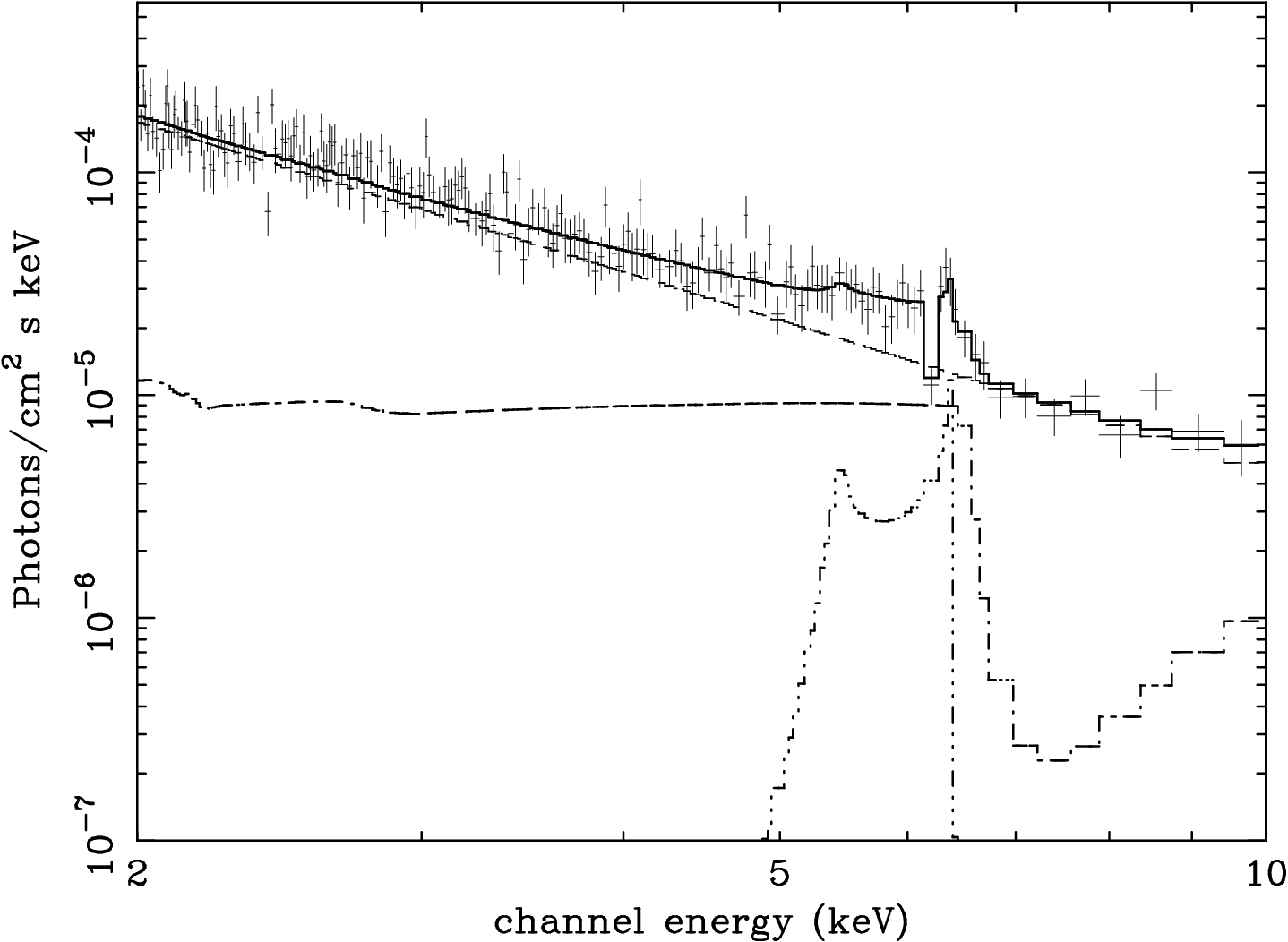,width=8.0cm,height=6cm,angle=-90}}
\  \hspace{.1truecm}     \
\parbox{8truecm}
{\psfig{file=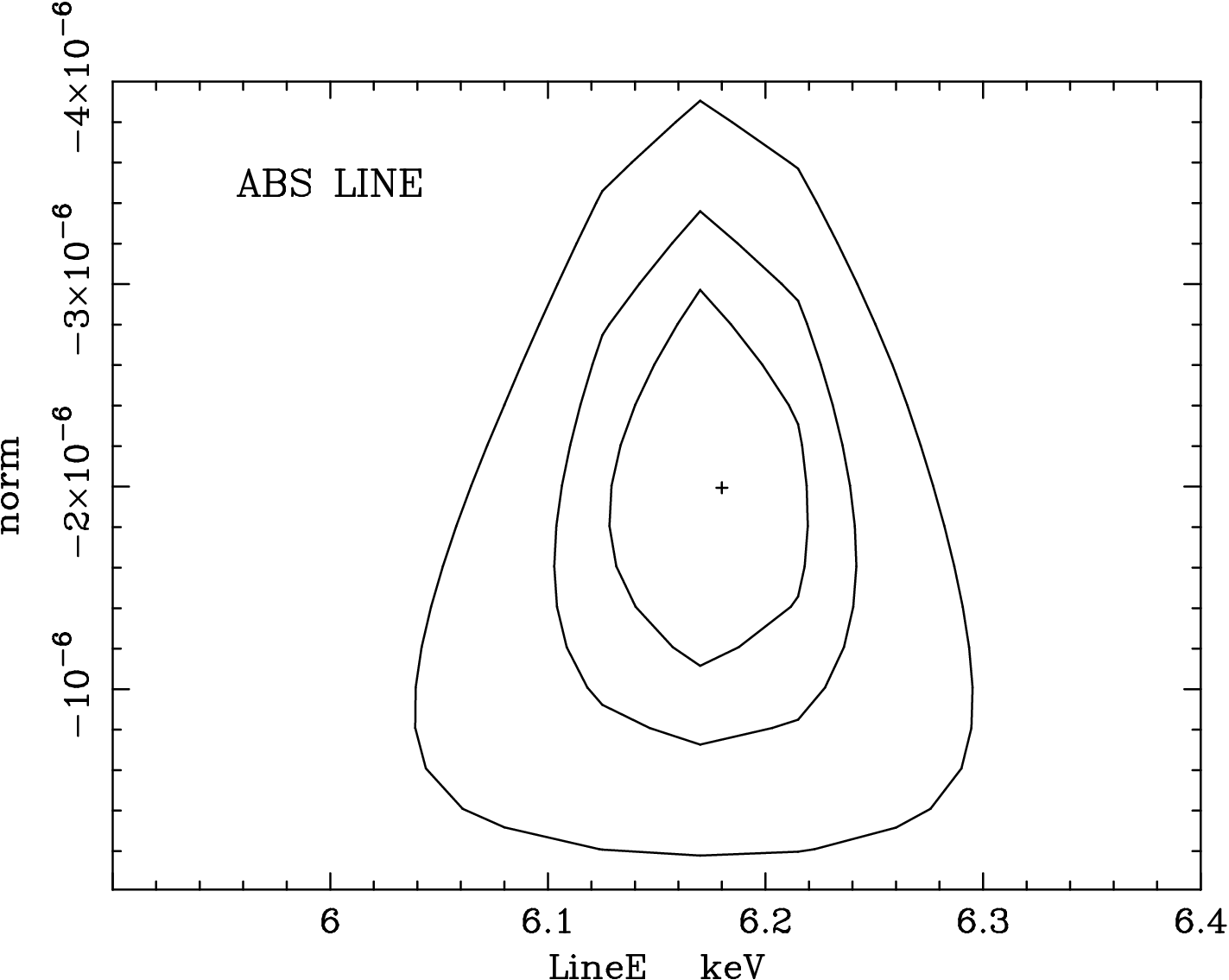,width=8.0cm,height=6cm,angle=-90}}
\parbox{8truecm}
{\caption{\small  Observer's frame picture of the unfolded spectrum for model 2 in Table 1. The rest frame energies are 6.84$^{+0.17}_{-0.15}$ keV for the Schwarschild diskline  and 
6.86$^{+0.08}_{-0.10}$ keV for the resonant absorption line superimposed on the previous one.}}
\  \hspace{.1truecm}   \
\parbox{8truecm}
{\caption{\small Observer's frame confidence contours at 68\%, 90\% and 99\% confidence 
level for the energy and the intensity  of the  narrow absorption line in Model 2: the absorption line is plotted  with negative intensity.}}
\end{figure*}    
Our investigations of various line models 
resulted in two possibilities, model 2 and model 3 in Table 1, respectively with disk 
line in Schwarschild metric  or  Kerr metric.  
Disk line  models provide a good fit to the red wing of the data; 
such  models consist in several parameters and so they are usually  affected by 
considerable degeneracy.
The model complexity does not allow to constrain all the parameters, but  
if the inner radius is left free to vary, it clearly tends to its minimum 
value (6 R$_g$) in Schwarzschild case.
With this low value, the Fe line emission is strongly 
concentrated  in the central regions, which suggests that the disk 
extends closer to the black hole where the relativistic effects are 
most important.
The Kerr metric allows to extend the emissivity region to the 
last stable orbit (1.23 R$_g$) and the inner radius reaches such low values that it 
has been fixed at its minimum  in order to avoid local minima for the $\chi^2$ process.
The same procedure has been adopted for the  emissivity law index (emissivity scales as 
R$^{-\beta}$)
which tends to $\beta$ $\sim$ 2 in both line models and it is kept fixed.
For the sake of simplicity, a separate table (Table 2) lists the relativistic line 
parameters of the best fits 2 and 3;  the outer radii and the disk inclination angles 
span a wide range of values so, again, we decided to fix them at the values most commonly
given by the fitting. 
 
However, the line profile cannot be well fitted with a single 
component, even considering relativistic effects for every spectral feature, as done 
so far; 
such models are inadequate to fit the ``sharp'' notch at E $\sim$ 6.2 keV 
(observer's frame, Fig. 1); we were thus forced to use double-component solutions, 
 listed in Table 1 and illustrated in Fig. 4 and 5.
A narrow Gaussian emission line was added  to the Laor disk line (model 3), 
obtaining a $\chi^2$ of 214/212 degrees of freedom (Fig. 4).
The rest frame energies of both the lines are very ionized, being 6.67$^{+0.72}_{-0.67}$ 
keV for the 
broad one and 7.02$^{+0.25}_{-0.06}$  keV for the narrow one. 
Then, the next double component model consists in a Schwarzschild disk 
line plus a narrow Gaussian absorption  line with $\sigma$ fixed at 0 keV and  negative 
intensity, in order to account for the narrow absorption notch, visible at 6.2 keV (observer's
frame) in Fig. 1 (zoom). The absorption line energy is 6.86$^{+0.07}_{-0.14}$ keV, which is just slightly 
different from the peak energy of the broad one, E=6.84$^{+0.29}_{-0.26}$ keV.
The confidence contours in Fig. 6 show that the absorption line is well constrained and it improves the fit by 
$\bigtriangleup$$\chi^2$=10  for two free parameters, i.e.$\ge$ 99\% of significance.  
This model provides the best fit of all the reflection ones, 
giving a $\chi^2$ = 211/212 and is the one shown in Fig. 5.

While we disfavour above the possibility that the Fe K-edge arises
from a fully or partially covering absorber, there is nonetheless
clear evidence in the RGS spectrum for ionized gas in line-of-sight
to IRAS 13349+2438 (Sako et al. 2001a). Such a component may affect our fits
if it has a sufficiently high column density. The highest column
component identified by Sako et al. has N$_H$$\sim$(1-4)$\times$10$^{22}$ cm$^{-2}$ 
with an  ionization parameter in the range 2.0 $\leq$ log $\xi$ $\leq$ 2.5. 
This is indeed high enough to have a small effect on
the spectrum around 2 keV and could change the parametrization of
the line. We have therefore added a completely covering ABSORI
component to the above fits. This provides a significant
improvement of the fit, yielding a $\chi^2$= 200/211 d.o.f.
The continuum photon index is much more steeper than previous models, 
being $\Gamma$=2.67$^{+0.20}_{-0.07}$, and the reflection fraction is R=1.13$^{+0.28}_{-0.27}$.
The power law is absorbed by a column of ionized gas N$_w$$\sim$ (2-5) $\times$10$^{22}$ cm$^{-2}$ and the 
ionization parameter of the absorbing medium is distributed in a range  
1 $\leq$ log$\xi$ $\leq$ 2, consistent with the value inferred by RGS data by Sako et al. (2001a).
The FeK edge optical depth is considerably lower than found before, 
($\tau$=2.07$^{+2.13}_{-1.07}$),
probably due to the higher photon index, while the energy is still 7.30$\pm$0.28 keV.
There is very little affect on the line parameters, however, and we
therefore conclude that the ionized absorption inferred by Sako et
al. (2001a) in this objects will not change our interpretation of the
 Fe-complex significantly.

\section{Discussion}
\subsection{On the hard X-ray continuum and strength of the Fe-K features}
The properties of IRAS13349+2438 EPIC spectrum are discussed, mainly 
concentrating on the Iron line region.
The X-ray spectrum turns out to be at least unusual, if not ambiguous.
Throughout the spectral analysis, the Fe K-edge is used as a key to
understand which is the underlying  physical model.
Its energy at $\sim$ 7.3-7.4 keV indicates a medium degree of ionization 
(Fe VIII-XII); from the spectral fitting a transmission  edge is ruled out,
 as the transmission models in sect. 3.1 clearly do not provide a good fit for the 5-7 keV 
residuals and also they require a very high Iron over abundance (see Fig. 3). 
Moreover, we would expect
an almost neutral Fe emission line  associated to the photoabsorption edge, 
produced by transmission through the medium along our line 
of sight: the Iron absorbs the primary radiation coming from the central source  and 
re-emits X-ray photons by fluorescence correspondently to its ionization 
state. But this is not consistent with  our case 
since we do not detect {\itshape any} neutral or near-neutral
emission line.

The next hypothesis of a reflection Iron edge has brought us to
use reflection models for the continuum, which can also fit
the 5-7 keV ``bump'' in the data with the associated FeK$\alpha$ emission.
Recent publications have stated the opportunity of using 
self-consistent ionized disk models
which can fit simultaneously the reflection continuum and the Iron features
(Nayakshin \& Kallman 2001; Ballantyne et al. 2001).

Nevertheless we found that the more
phenomenological PEXRIV/SMEDGE model gave a better fit, due to the
complexity
of the Fe line/edge region and the consequent necessity of an 
{\itshape ad hoc},
(and hence flexible) model\footnote{Using
diskline models as we did, allows a greater control on the different 
components and on the line
profile shape, which is affected by changing all the parameters in Table 2.}.
The photon index derived after adding the reflection component, 
($\Gamma$$\sim$ 2.2), is consistent with
the one found by ASCA  (Brandt et al. 1997b) and 
with the mean value of
$\Gamma$ estimated by Leighly (1999) for 23 NLS1.
As expected, once the spectrum is fitted with the reflection 
continuum, the power law slope
results steeper than it is in Model 1 by $\bigtriangleup$$\Gamma$ 
$\sim$0.18  (Nandra \& Pounds, 1994).
The reflection fraction  includes the uncertain information 
concerning the geometry
of source and
reflector.
We find R=1-2, depending on the continuum parametrization.
This strong reflection is primarily driven by the deep edge in the data. 
It is consistent with the high equivalent width of the
line (EW $\sim$ 530 eV), with the latter also being enhanced by the 
high ionization state
implied by the line energy.
High reflection fractions
have already been found in Seyfert 1 galaxies and cannot be excluded 
{\itshape a priori} (Matt et al.  2003).
Nonetheless it is troubling that we
have been unable to find a self-consistent physical model which
simultaneously account for both the line and the deep edge.

In this regard, an interesting possibility
has been suggested by Fabian et al. (2002). In their
model
the spectrum is
reflection-dominated as a result of multiple reflections from 
different layers of cold material, likely to be produced in 
high-accretion rate systems as NLS1 as a consequence of accretion 
instabilities. They
find a good fit to another NLS1 with a deep edge, 1H0707-49 (Boller
et al. 2002).
We attempted to fit the broadband X-ray spectrum of IRAS13349+2438 
with such model using a combination of three reflectors plus a 
blackbody component of kT$\sim$76 eV to account for the thermal soft 
X-ray emission from the disk. Each reflector is modeled using a 
constant-density ionized disk model (Ross \& Fabian, 1993; 
Ballantyne et al. 2001) on which relativistic blurring 
is applied appropriate to a maximally spinning black hole. A narrow 
Gaussian emission line at $\sim$ 7 keV rest frame  is also included 
in the fit. This model is not only able to account for the line and 
edge features in IRAS13349+2438, but provides a quite  reasonable fit to the 
entire broadband spectrum (0.3-10 keV; $\chi^2$=675/545  d.o.f);
The three reflectors have all different ionization parameters indicating
the variously ionized  layers in the disk: log$\xi$=1.977, with 
R$_{in}$=2.08 R$_g$,  log$\xi$=3.957 with R$_{in}$=13 R$_g$, 
log$\xi$=1.494 with R$_{in}$=23 R$_g$  going towards the outer disk. 
This fascinating
possibility merits further exploration with future data.

\subsection{On the FeK line structure}     
The FeK$\alpha$ diskline is surely the most interesting feature in the 2-10 keV spectrum of 
IRAS13349+2438.   
It is clearly well modelled by the superposition of two components, 
for which two scenarios have been found:
i) the sum of two emission lines or ii) a single, double-peaked diskline 
plus a narrow absorption line.
Disk models give an excellent fit to the broad red wing and strongly indicate 
 emission coming from the accretion disk inner regions.
From these data we cannot distinguish between rotating and nonrotating black 
hole models, although a Kerr geometry seems more likely as it allows the 
emitting region extending very close to the black hole.
In every model, the peak rest energy (6.7-6.9 keV) implies the presence of 
Iron atoms at a very high ionization level (Fe XXIII-XXVI) in the 
accretion disk.
Such high states can account for the high EW of the line, as there is a considerable 
enhancement of the fluorescence mechanism corresponding to Fe XX-Fe XXVI 
(Zycki \& Czerny, 1994) and also a ionized reflection is likely to produce such a strong line for particular source geometries (Matt et al. 1993).
 
In case i) the broad hump at 5-6 keV in Fig. 1, is fitted by an emission diskline in Kerr metric,
whereas the narrow peak at $\sim$ 6.3 keV in the same figure, is accounted by a second
emission line, which is very ionized. 
The current interpretation for a narrow Fe K emission line as being due to 
fluorescence from  material distant from the inner disk region, seems the 
most appropriate.
One possibility of producing the emitted Iron line is reflection onto optically 
thin material, far from the central region, for which  the so-called warm mirror 
might be a candidate (Krolik \& Kallman, 1987).
Improved measurements of the line width and flux variability will better 
constrain the origin of the narrow Fe K$\alpha$.  

In the ii) hypothesis, the red hump plus a narrow emission peak are  
interpreted as a single Schwarzschild diskline and the narrow absorption component 
 at 6.8 keV that is produced by resonant absorption;
this is a scattering process caused  by Hydrogen- and Helium-like Fe ions 
which resonantly absorb the fluorescent Iron line and  continuum  photons. 
The resonant transitions are expected to occur  at the energies of 6.7 keV and 
6.9 keV. 
The total Iron line profile is then influenced by the absorption feature.

Revealing and studying with greater details such resonant absorption
lines is a key point in determining not only the FeK$\alpha$
line profile, but also the  accretion disk environment.
In fact, the first case was a very
redshifted  absorption feature found in the Seyfert 1
NGC 3516 by Nandra et al. (1999), which represents a tentative and 
rare direct evidence for
material actually accreting  onto the black hole in active galactic nuclei.
Theoretical calculations of the resonant absorption line EW are 
available in Matt et al. (1997).
It is also of note that resonant absorption features in the Fe K region are now
becoming detectable even with low resolution  in  XMM-Newton EPIC 
spectra (Pounds et al. 2003); when redshifted, they might be
observable in  brighter objects with higher precision with long 
exposure of  Chandra gratings.
The IRAS 13349+2438 resonant line, although not redshifted, shows the 
existence of a plasma of Fe ions placed somewhere above the corona, 
capable of imprinting
significant absorption features on AGN spectra.
This gas is not to be confused with the soft X-ray warm absorbers 
found by Sako et al. which are much less ionized; it might be
a signature of a thin skin or wind sourrounding the accretion disk 
(Ruszkowski \& Fabian, 1999).
Variability studies of the line
profile could be a powerful tool for constraining the size and position of
this component and accurate measurements of the absorption line width could
  provide valuable information on the velocity gradient  within
the gas. 
For meaningful further progress, future observations need 
both high resolution and good signal-to-noise ratio. Observations with
Astro-E 2 and later Constellation-X are therefore best suited to discriminate
between the possible scenarios for the origin of the Fe line complex 
in this galaxy.

\begin{acknowledgements}
We thank the XMM-Newton Calibration Team for contributing to the operations of the satellite and for continuous maintenance of the software and the anonymous referee who contributed to improve this paper.
Financial support from the POE network at the Imperial College, London, and from Italian Space Agency by contract N.ASI/I/R/107/00 is acknowledged.
The authors wish to thank Andy Fabian for providing the relativistic 
blurring codes. 
A.L. Longinotti is pleased to thank Cristian Vignali for his constructive comments and 
suggestions.

\end{acknowledgements}

\begin{figure*} [tb] 
\begin{center} 
\vspace{-1cm}
\hspace{12cm}
\psfig{figure=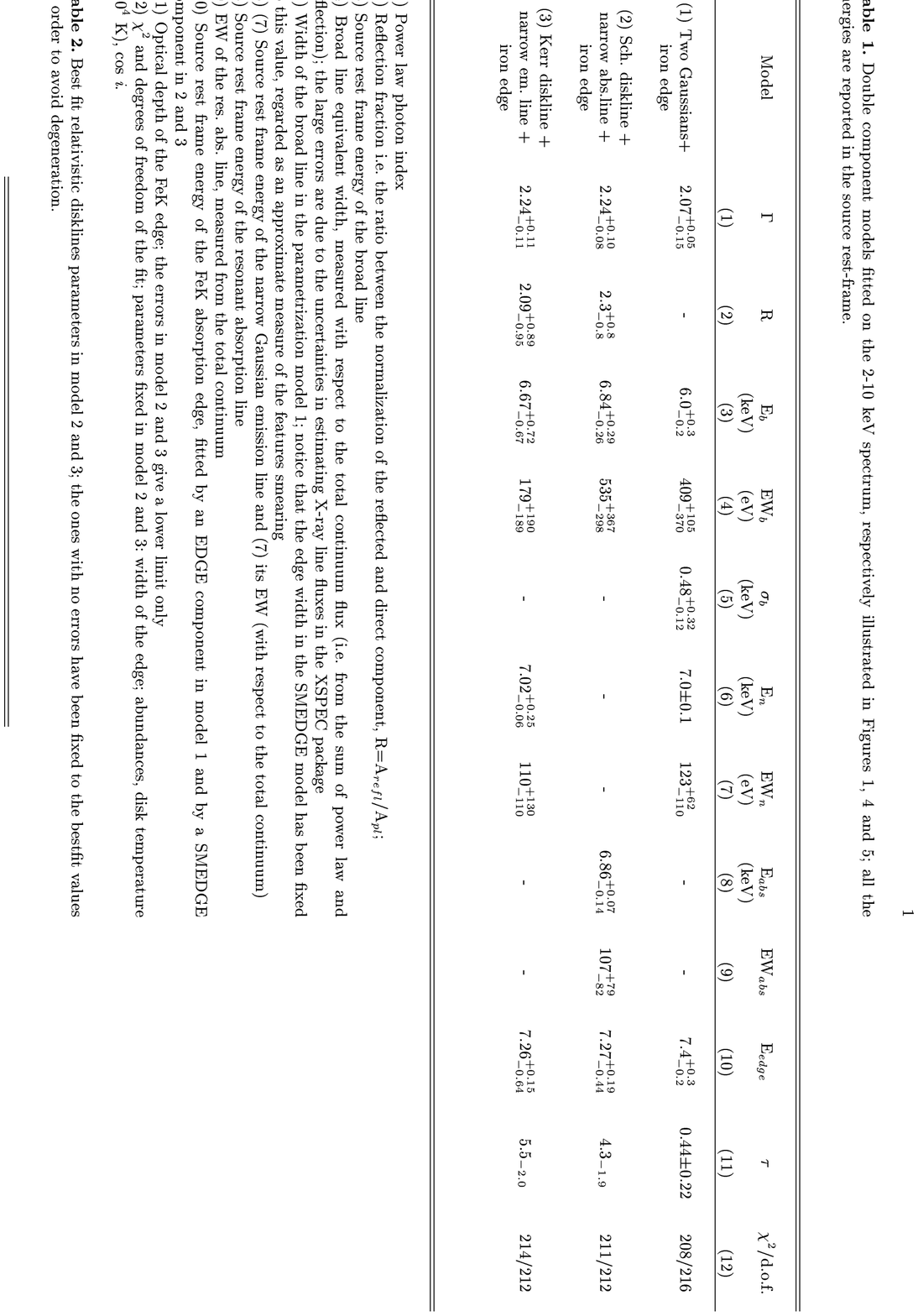,height=25cm,width=15cm,angle=0}
%\caption{}
\end{center} 
\end{figure*}

\vfill\eject
\onecolumn

\end{document}